\documentclass[doublecol]{epl2}

\title{Magnetic screening in proximity effect Josephson-junction arrays}

\author{Mauro Tesei\inst{1} \and Ricardo Th\'{e}ron\inst{2}}

\institute{
  \inst{1} The Blackett Laboratory, Imperial College - London SW7 2BZ, UK, EU\\
  \inst{2} Ecole Polytechnique F\'{e}d\'{e}rale de Lausanne (EPFL), Institute of Microengineering IMT, Photovoltaics and thin film electronics laboratory - Neuch\^{a}tel,  Switzerland}

\pacs{74.81.Fa}{Josephson junction arrays and wire networks}

\pacs{74.25.Ha}{Magnetic properties}

\pacs{64.60.De}{Statistical mechanics of model systems}

\abstract{The modulation with magnetic field of the sheet inductance measured on
proximity effect Josephson-junction arrays (JJAs) is progressively vanishing on lowering the
temperature, leading to a low temperature field-independent response. This behaviour is consistent
with the decrease of the two-dimensional penetration length below the lattice parameter. Low
temperature data are quantitatively compared with theoretical predictions based on the XY model in
absence of thermal fluctuations. The results show that the description of a JJA within the XY
model is incomplete and the system is put well beyond the weak screening limit which is usually
assumed in order to invoke the well known frustrated XY model describing classical Josephson-junction arrays.}

\begin{document}

\maketitle

\section{Introduction}
Proximity effect Josephson-junction arrays (JJAs) exposed to a perpendicular magnetic field $B$ are
well known to be, under some conditions discussed below, a good physical realization of the
frustrated XY model ~\cite{XY} and so have been studied for many years as model systems, and still
reveal some interesting physics ~\cite{Review}. The description of JJAs in the framework of the
frustrated XY model remains valid in a broad temperature range below the superconducting-to-normal
transition temperature $T_{c}$. In the case of proximity effect SNS JJAs, quantum
fluctuations are irrelevant even at very low temperature. Indeed, the Josephson coupling energy
$E_{J}=\phi_{0}I_{c}/2\pi$ ($\phi_{0}$ the flux quantum and $I_{c}$ the single junction critical
current) is much larger than the Coulomb charging energy $E_{c}=e^{\ast2}/2C$ ($e^{\ast}=2e$ the
free Cooper pair charge and $C$ the junction capacitance), and the normal state resistance is much
smaller than the quantum resistance ~\cite{Fazio_SCNetworks} (roughly 7 orders of magnitude smaller
in our systems). These two criteria ensure that our JJAs remain in the classical regime for all
$T<T_{c}$. However, when lowering the temperature magnetic interactions (between screening currents of
neighbouring cells) put the system beyond the pure XY model. As a matter of fact, the system
experiences a gradual crossover from a \emph{high temperature XY weak screening regime} where
vortex ground states satisfy the fluxoid quantisation in all plaquettes ~\cite{XY}, and local
magnetic field equals the applied field $B$, to a \emph{low temperature strong screening regime}
with quantised flux where $B$ penetrates the JJA in a similar way to how it penetrates a
superconducting wire network creating an Abrikosov-like lattice. In the weak screening
limit, the frustration parameter $f$ is defined by the magnetic flux $\phi$ threading an elementary
cell in units of the flux quantum $\phi_{0}$; $f=\phi/\phi_{0}$. While lowering the temperature, the JJA is gradually put beyond the weak screening limit when supercurrents flowing
around each lattice cell are no longer negligible. At very low temperatures, when an elementary
cell of the array is able to fully screen a flux quantum, any connection to the XY model
is lost since the JJA behaves as a multiply-connected superconductor.\\
Although their existence is well known, the effects of screening currents on the behaviour of JJAs
have been investigated mainly theoretically, in the weak screening regime (as a correction of the
frustrated XY model) ~\cite{weak_screening} and beyond ~\cite{RSJ_size,strong_screening}. This paper focuses on the low temperature regime of JJAs on a periodic lattice with
hexagonal symmetry called a \emph{dice} lattice (see Fig. 1 and description below) where the
measured sheet inductance is independent of $B$. In this regime, three different phenomena push our JJA beyond the XY model. Firstly, the effective penetration length becomes shorter than the lattice parameter and the magnetic energy associated with screening currents dominates over the Josephson coupling energy as far as the magnetic field response is concerned. Secondly, thermal fluctuations gradually vanish and vortex mobility is accordingly reduced. Finally, on a smaller size scale, the Josephson penetration length becomes smaller than the size of a Josephson junction which results in a non-unique phase difference across the junction width. All this leads to a low temperature regime dominated by magnetic screening effects which efficiently suppress the field modulation of the measured sheet inductance. This scenario is confirmed by comparing low temperature sheet inductance data with the same quantity calculated within the frustrated XY model in absence of thermal fluctuations,
\textit{i.e.} in the ground states. Thus, when the temperature is lowered JJAs can easily reach a
regime where the XY model is not sufficient to describe their behaviour.\\
In addition, for the first time data taken using the two-coil mutual inductance technique
~\cite{twocoil}, as well as the related data treatment, are confirmed by comparing $I_{c}$
extracted from inductive measurements with the same quantity measured by a conventional four probe
measurement technique.

\section{Measurement technique}

\begin{figure}
\onefigure{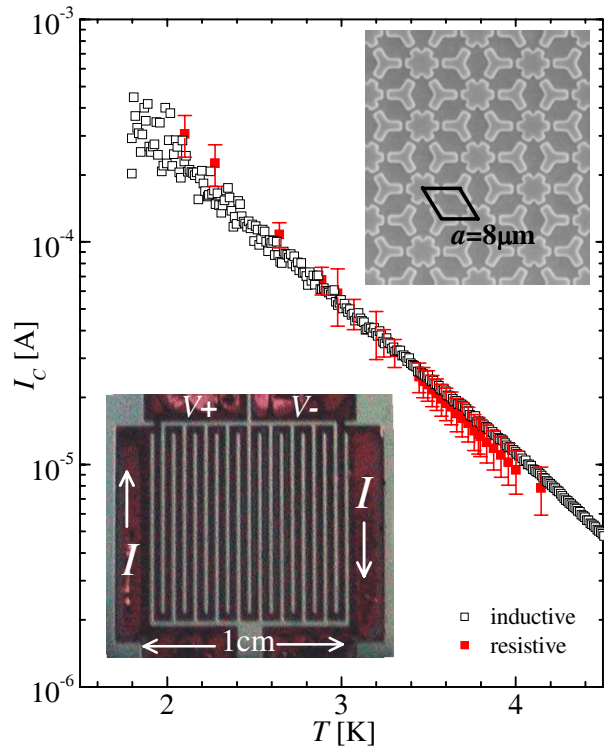} \caption{(colour on-line) Low temperature dependence of the single
junction critical current $I_{c}$ extracted from resistive measurements (red plain symbols) and
measured inductively (black empty symbols) at zero field. Upper inset: SEM picture of a portion of
a $1\times1 $cm$^{2}$ JJA on \emph{dice} lattice. The array consists of $\sim 10^{6}$ star-shaped
(6-fold and 3-fold) superconducting lead islands coupled by proximity effect junctions in the
underlying normal copper layer. The junctions have nominal width $w=2\mu$m and length $l=1\mu$m. An
elementary rhombic cell illustrated with black line has a side $a=8\mu$m. Lower inset: photograph
taken with an optical microscope of modified sample for resistive $dV/dI$ measurements. The current
$I$ is injected on copper contacts at bottom into the central zigzag strip, and voltage
$V+/V-$ is measured on copper contacts at the top ends of the strip. The strip consists
of 18 branches of width about $300\mu m$ containing 48 parallel junctions. The branches are equally
spaced by $100\mu m$ and go back-and-forth across an area slightly smaller than the initial JJA.
The strip is about 12cm long.} \label{fig.IcSample}
\end{figure}

Arrays of proximity effect Pb-Cu-Pb junctions on a \emph{dice} lattice (see upper inset to
fig.~\ref{fig.IcSample}) were probed using a SQUID-operated two-coil mutual inductance technique
~\cite{twocoil} that allows for measurements of the screening properties of superconducting wire
networks ~\cite{SWNs}, JJAs ~\cite{Review}, as well as high temperature superconducting thin films
~\cite{HTS}. Here the macroscopic measured quantity is the inverse sheet inductance
$L^{-1}$ (or inverse inductance "per square" ~\cite{Lobb_weaklinks}) that is inferred from the
array's linear sheet conductance ~\cite{twocoil}, and measures the degree of superconducting phase
coherence in the sample. The inductive technique has a sensitivity threshold of the
order of 10pH.

For comparisons with theoretical predictions based on the XY model and with samples with
different coupling constants, \textit{i.e.} different geometrical parameters, the temperature is
scaled according to the thermal energy $k_{B}T$ and $E_{J}$; the so-called \emph{reduced
temperature} is $\tau=k_{B}T/E_{J}$. At zero field and at temperatures well below $T_{c}$ the
sheet inductance of a regular array $L\propto L_{J}$ ~\cite{Lobb_weaklinks} the single junction
inductance, and the numerical factor depends only on the lattice geometry. For the \emph{dice}
lattice ~\cite{Tesei_thesis} $L^{-1}(T)=(2/\sqrt{3})L_{J}^{-1}(T)$ and
$L^{-1}(T)=(2/\sqrt{3})(2\pi/\phi_{0})I_{c}(T)$ ~\cite{Tinkham}. Close to $T_{c}$, the critical
current $I_{c}$ is extrapolated from low temperature data ~\cite{Zaikin}.\\
For the first time inductive measurements performed using the two-coil mutual inductance technique
~\cite{twocoil}, and the successive data treatment, are independently and directly confirmed by
comparing the temperature dependence of the single junction critical current $I_{c}$ (at zero
field) extracted from inductive measurements on a JJA with the same quantity measured by four probe
measurements following the procedure detailed in ref.~\cite{Ic_measurements} where $I_{c}$ is
defined as the current that produces a maximum in the dynamic resistance ($dV/dI$ vs $I$). Since
the small normal state sheet resistance of proximity effect JJAs (of the order of a
m$\Omega$) and the high current that one would have to apply to approx. 1000 parallel channels of
junctions in the 1cm$^{2}$ sample, and leading to a noticeable increase of temperature in the
array, the sample geometry was modified after all inductive measurements were done. A new
pattern is obtained by an additional optical lithography process with the photoresist layer spread
out on the top of the pre-existing JJA, and part of the superconducting lead islands were removed
using Ar-ion milling. The resulting pattern was a thin (48 parallel junctions) and 12cm long back-and-forth strip across the initial array, as shown in the lower inset of
fig.~\ref{fig.IcSample}. The pre-existing copper layer all around the JJA is left on the
substrate and used for current and voltage contacts, as well as to dissipate the heat due to the
measurement current (not shown in fig.~\ref{fig.IcSample}). This new pattern allows for a lower
applied current and the measured voltage drop across the strip is much greater than it would be
across the initial array. Particular care during the modification process was taken to avoid
residual lead islands in-between the branches of the strip that could create electric shortcuts
between the branches. Fig.~\ref{fig.IcSample} shows low temperature values of $I_{c}$ obtained
from both inductive measurements on the original JJA and resistive measurements on the modified
sample. The error in $I_{c}$ obtained by resistive measurements is due to the resolution in
measuring the current and the related position of the maximum in $dV/dI$. The measurements are
limited to temperatures below 4.2K because the sample was immersed in the helium bath of the
cryostat to avoid local heating of the junctions which could be induced by the applied current. Insight into the effects of the reduction in the size of the array can be provided by a
comparison of our experimental conditions with previous results obtained by numerical simulations
of the resistively shunted Josephson-junction model at zero temperature ($T=0$) ~\cite{RSJ_size},
thus neglecting finite temperature effects. At zero field and at $T=0$ the dynamics of a JJA
reduces to that of a set of uncoupled channels of single junctions along the current direction
~\cite{RSJ_size}. In our experiments the ambient magnetic field is reduced by a combination of mu-metal and
superconducting screens and at nominally zero field the residual frustration is $f=10^{-3}$
($B\cong0.4$mG), which we can reasonably assume small enough to compare our data with results at
$f=0$ ~\cite{RSJ_size}. A frustration parameter $f=1$, \textit{i.e.} one quantum of flux per cell,
is achieved with a perpendicular magnetic field $B\cong 360$mG. When edge fields induced by the
applied external current are taken into account and in the case of a 32x32 array [4] (compared to
our modified sample with 48 parallel junctions), $I_{c}$ is reduced by less than 10\% when the
field penetration length is of the order of the lattice parameter, \textit{i.e.} in the
strong screening regime (see discussion below). In the following we show that in our JJAs such a
strong screening regime is achieved for temperatures $T<3$K. Thus we do not expect our resistive
measurements to be significantly affected by the reduced size of the sample, except possibly for
the few measurements below 3K in fig.~\ref{fig.IcSample}. The data presented in
Fig.~\ref{fig.IcSample} show the validity of the measurement technique, as well as the procedure to
extract the critical current.

\section{Results and discussion}

\begin{figure}
\onefigure{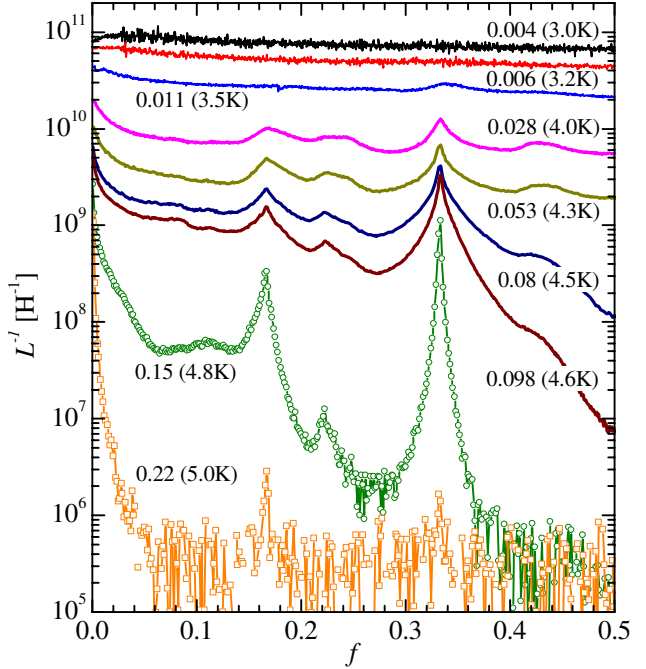} \caption{(colour on-line) Inverse magnetoinductance $L^{-1}(f)$ isotherms
measured in the temperature range between $\tau=0.22$ (T=5K) and $\tau=0.004$ (T=3K) with an
excitation frequency $\omega/2\pi$=16Hz.} \label{fig.Lvsf}
\end{figure}

In the weak screening regime, due to the nature of the coupling in JJAs ~\cite{XY}, the frustration
$f$ can induce a pronounced modulation of the single junction inductance
$L_{J}(T)=(\phi_{0}/2\pi)/(I_{c}(T)cos(\theta))$ ~\cite{Korshunov_magnetoinductance}, hence the
measured sheet inductance $L(T)$, where $\theta$ is the gauge invariant phase difference related to
$f$ by the fluxoid quantisation ~\cite{XY,Review}. Fig.~\ref{fig.Lvsf} shows the response of a JJA
while sweeping $B$ in the reduced interval $0<f<1/2$ ~\cite{XY}.

Well defined structures are measured at low excitation frequency showing peaks in $L^{-1}(f)$ which
reveal enhanced superconducting phase coherence for some rational values of frustration
($f=0,1/6,2/9,1/3$). The $f=1/3$ state has been the subject of previous work which shows that a
strongly enhanced superconducting phase coherence, as for $f=1/3$, does not necessarily imply that the vortex pattern is ordered ~\cite{dice-tiers.serret,dice-tiers.korshunov},
even if ordering in the vortex pattern is normally associated with the commensurability
between the vortex lattice induced by $B$ and the array geometry, and resulting in a strong phase
coherence as shown in fig.~\ref{fig.Lvsf} for $f=1/3$. The fully frustrated state ($f=1/2$), which
on the contrary shows a pronounced depressed response in the intermediate temperature range, was
also investigated ~\cite{dice-demi.exp,dice-demi.theory} and for both $f=1/3$ and $f=1/2$ states
the magnetic interaction between screening currents circulating in neighbouring cells of the array
has been considered, but still assuming a uniform frustration over the array. Starting from the
hottest isotherms and lowering the temperature, $L^{-1}(f)$ first shows the appearance of well
defined structures at $f=0,1/6,1/3$ which grow up until some temperature $\tau\approx0.1$. While
further lowering the temperature the structures progressively disappear as shown by the coldest
isotherms in fig.~\ref{fig.Lvsf}.\\

\begin{figure}
\onefigure{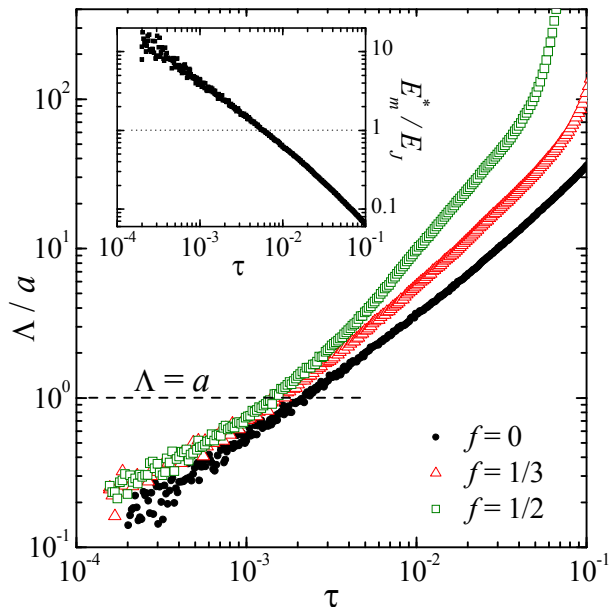} \caption{(colour on-line) Low temperature dependence of the effective
penetration length $\Lambda$ in unit of the lattice parameter $a=8\mu$m (see upper inset
to Fig.~\ref{fig.IcSample}) for three frustration values $f=0,1/3,1/2$. Inset: energy ratio
$E_{m}^{\star}/E_{J}$ as a function of reduced temperature $\tau$.} \label{fig.Lambda}
\end{figure}

Clues about the very low temperature response ($\tau<10^{-2}$) come from the temperature
behaviour of the array penetration length. For a thin superconducting film of thickness $d$ and
exposed to $B$ the screening distance is given by a thickness-dependent penetration length
~\cite{Pearl_lambda}, the bulk material (3D) penetration depth $\lambda$ being rescaled according
to the thickness $d$, the effective 2D penetration length is $\Lambda=2\lambda^{2}/d$.
In the case of an array, $\Lambda$ can be expressed in terms of $I_{c}$ ~\cite{Tinkham}, or as
mentioned above as a function of the measured sheet inductance; $\Lambda=2(2/\sqrt{3})L/\mu_{0}$
for a \emph{dice} lattice. Fig.~\ref{fig.Lambda} shows the low temperature dependence of
$\Lambda$ expressed in unit of the lattice parameter $a$ for three values of frustration: $f=0$,
$f=1/3$, and $f=1/2$.

\begin{figure}
\onefigure{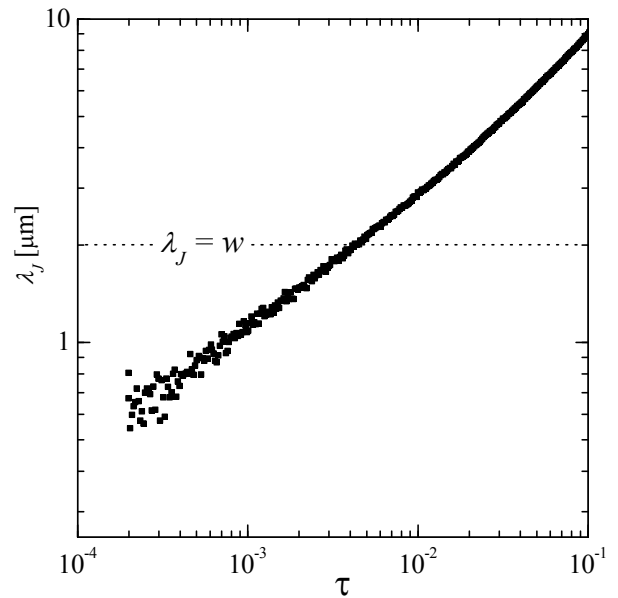} \caption{Josephson penetration length
$\lambda_{J}$ as a function of reduced temperature $\tau$. The dashed line indicates the size of a
juntion (width $w=2\mu$m).} \label{fig.Em}
\end{figure}

At very low temperature, below $\tau\approx 3\cdot 10^{-3}$, the field dependence of $\Lambda$
(\textit{i.e.} the superconducting phase coherence) is suppressed in good agreement with the
magnetoinductance isotherms of fig.~\ref{fig.Lvsf}. Most interestingly, this temperature is very
close to the temperature at which $\Lambda=a$, the lattice parameter. In other words, for $\Lambda$
close to or smaller than the lattice parameter the sample no longer behaves as a JJA with
characteristic size $a$. This observation suggests the presence of strong magnetic screening. More
information is obtained by comparing the magnetic energy $E_{m}$ stored in each supercurrent
$I_{s}$ loop around the lattice plaquettes with the Josephson energy $E_{J}$. For the sake of
simplicity $E_{m}$ is overestimated by considering the maximum supercurrent, \textit{i.e.} the
critical current $I_{c}$; $E_{m}=\frac{1}{2}\mathcal{L}I_{s}^{2} \leq
\frac{1}{2}\mathcal{L}I_{c}^{2}=E_{m}^{\ast}$. The geometrical inductance $\mathcal{L}=26$pH of a
rhombic loop (see upper inset to fig.~\ref{fig.IcSample}) is calculated using the numerical results
of ~\cite{Grover_inductances} for a wire with rectangular cross section. The ratio
$E_{m}^{\ast}/E_{J}$ shown in the inset of fig.~\ref{fig.Lambda} shows that the magnetic energy is
dominating over the Josephson energy for temperatures $\tau<5\cdot 10^{-3}$.\\
Screening effects also appear locally on each Josephson junction, inside which Meissner screening fields are responsible
for inhomogeneous distribution of the Josephson currents. Quantitatively this phenomenon is
negligible for small junctions, \textit{i.e.} when the junction width $w \ll \lambda_{J}$ the
Josephson penetration length given by $\lambda_{J}=\sqrt{\hbar/(2e\mu_{0}lJ_{c})}$ ~\cite{Barone},
where $J_{c}$ is the cross section critical current density and $l$ is the junction length (see
fig.~\ref{fig.IcSample}). Fig.~\ref{fig.Em} shows the temperature dependence of $\lambda_{J}$ which
is smaller than the junction width $w$ for $\tau<4\cdot 10^{-3}$. Thus, at lower temperature the current distribution inside the junctions is no longer homogeneous. As a consequence, the phase difference across the junction width is no longer unique and the JJA cannot be mapped onto the XY model.\\

There exist two different mechanisms for the suppression of the magnetic field dependence
of the sheet inductance measured at finite frequency while the temperature is lowered: magnetic
screening induced by superconducting currents in the array and the vanishing of thermal
fluctuations leading to the freezing of the vortex diffusion between array cells. Evidence for the
predominance of the first mechanism at the lowest temperatures comes from a comparison of the
measured sheet inductance with the same quantity calculated in the XY regime and in absence of
thermal fluctuations, \textit{i.e.} the ground state value of $L^{-1}(f)$ calculated within the
frustrated XY model at selected frustration states $f=0,1/3,1/2$. At very low temperature where
thermal fluctuations can be neglected, the JJA is equivalent to a network of inductances
~\cite{Yu_LGamma} $L_{ij}(q_{ij})$ and the phase variables $q_{ij}$ are known in the frustrated
ground states at $f=1/3$ ~\cite{dice-tiers.korshunov} and at $f=1/2$ ~\cite{dice-demi.theory}. Each
ground state is characterised by a sheet inductance $L(f)$ that can be calculated from the
frustrated inductance network $L_{ij}$ which is first transformed into an anisotropic triangular
lattice $L'_{ij}$ by the triangle-star transformation ~\cite{Korshunov_magnetoinductance}. $L(f)$
is then calculated applying Kirchhoff's laws to the elementary cell of the transformed network
$L'_{ij}$.

\subsection{Unfrustrated state $f=0$}
All bonds of the transformed triangular lattice have the same inductance value, hence the isotropic
sheet inductance of the dice lattice $L(0)=(\sqrt{3}/2)L_{J}$ with $L_{J}$ the single junction
inductance at zero field.

\subsection{Frustrated state $f=1/3$}
Here we consider the (honeycomb) ground state which is dominating on a short scale over the
possible vortex configurations ~\cite{dice-tiers.korshunov}. The transformed sheet inductance is
isotropic; $L(1/3)=(4\sqrt{3}/5)L_{J}\cong1.39L_{J}$.

\subsection{Fully frustrated state $f=1/2$}
Applying the triangle-star transformation to the four periodic ground states
~\cite{dice-demi.theory}, we obtain two different configurations each with two inductance
components. Thus, in the ground state of the fully frustrated XY model the JJA behaves as a two
component inductance network ($L_{x},L_{y}$). Moreover, due to the inefficiency of the degeneracy
removal mechanism and the prominence of the finite size effects ~\cite{dice-demi.theory} the vortex
pattern can be assumed as disordered. In a 2D disordered system with two phases equally
distributed, the conductivity is given by the geometric mean of both phase conductivities
~\cite{Dykhne_disorderSystem}. Although in principle one should then consider the geometric mean of
($L_{x},L_{y}$), the discussion on the choice of the mean value, \textit{i.e.} the arithmetic or
the geometric mean, is irrelevant since they are almost the same and the experimental resolution
does not allow us to distinguish the two. The four periodic states ~\cite{dice-demi.theory} share the same geometric mean
$L(1/2)=3\sqrt{3}/(2\sqrt{2})L_{J}\cong1.84L_{J}$.\\

\begin{figure}
\onefigure{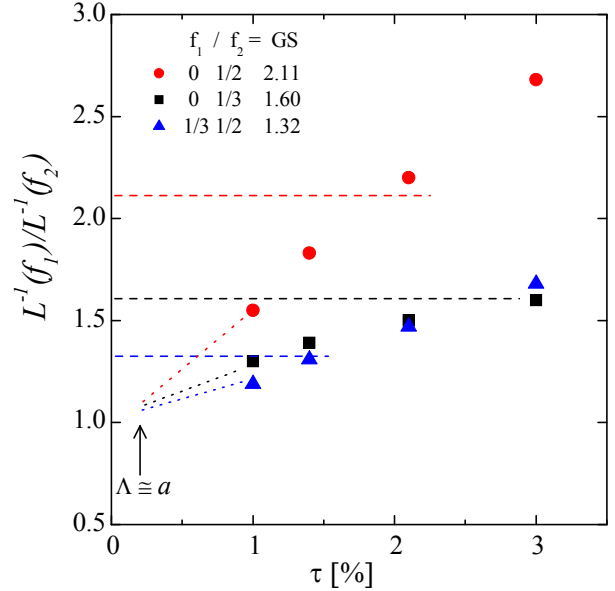} \caption{(colour on-line) Low temperature sheet inductance ratios
$L^{-1}(f_{1})/L^{-1}(f_{2})$ at selected frustration states $f_{1},f_{2}=0,1/3,1/2$ extracted from
high frequency ($\omega/2\pi=16$kHz) magnetoinductance isotherms. Horizontal (coloured) dashed
lines show the values of the ratios calculated in the corresponding frustrated XY ground states
(GS). The arrow indicates the temperature at which $\Lambda\cong a$ (extracted from
fig.\ref{fig.Lambda})} \label{fig.LRatios}
\end{figure}

The calculated sheet inductances $L(f)$ are theoretical predictions based on the XY model in
absence of thermal fluctuations and so have to be compared with low temperature data taken at high
frequency where the corresponding time scale is too short to allow for fluctuations. Figure
\ref{fig.LRatios} shows ratios of $L^{-1}(f)$ extracted from low temperature inverse
magnetoinductance isotherms like those in fig.\ref{fig.Lvsf} but at much higher frequency
($\omega/2\pi=16$kHz) and at $f=0,1/3,1/2$.
The data in fig.\ref{fig.LRatios} unambiguously demonstrate that at low temperature the modulation
of the measured sheet inductance is weaker than it would be in the related XY model in absence of
thermal fluctuations, and the selected ratios tend to unity. This behaviour can definitively be
ascribed to magnetic screening effects.

\section{Conclusion}

The field modulation of the sheet inductance in Josephson-junction arrays on a \emph{dice} lattice
is shown to vanish while lowering the temperature. The response is field-independent below a
reduced temperature $\tau=k_{B}T/E_{J}\approx 10^{-3}$, or equivalently below an easily accessible
temperature $T\approx$ 3K. Note that this threshold temperature may change in samples with
substantially different geometrical parameters. The very low temperature regime is characterised by
a penetration length shorter than the lattice parameter, and the magnetic energy associated with
screening currents is greater than the Josephson coupling energy. The observed behaviour is
explained in terms of increasing magnetic screening, with decreasing temperature, until a
strong screening regime is reached where the modulation of the measured sheet
inductance is weaker than in the frustrated XY model in absence of thermal fluctuations. The
disappearance of the modulation in the magnetoinductance at low temperature could also be
attributed to vortices with very low mobility. However, this would give rise to a hysteretic
response that was not observed down to the lowest accessible temperatures. In the (very) low
temperature screening regime, the JJA behaves as a multiply connected superconductor, and the XY
model is no longer valid. Unlike the weak screening regime where vortices are phase configurations
satisfying the fluxoid quantisation, in this strong screening regime vortices are real magnetic
objects carrying an integer number of flux quanta and interacting with currents through Lorentz
forces. These strong effects highlight the importance of a careful interpretation of low
temperature results within the frustrated XY model.

\acknowledgments MT is grateful to V. I. Marconi for the critical reading of the manuscript.
The data presented in this paper were taken when both authors were working in the group of P.
Martinoli at the University of Neuch\^{a}tel (Switzerland) and we appreciated his comments and
suggestions. This work was supported by the Swiss National Science Foundation.

\end{document}